\documentclass{PoS}

\usepackage{graphicx}
\usepackage{color}

\def \be  {\begin{equation}}
\def \ee  {\end{equation}}
\def \ee  {\end{equation}}
\def \bea {\begin{eqnarray}}
\def \eea {\end{eqnarray}}

\renewcommand{\figurename}{{\bf Fig.}}
\renewcommand{\tablename}{{\bf Tab.}}

\title{Baryon-to-pion ratios within generic (non)extensive statistics}

\ShortTitle{}

\author{\speaker{Abdel Nasser Tawfik}\\
Egyptian Center for Theoretical Physics (ECTP), Modern University for Technology and Information (MTI), 11571 Cairo, Egypt\\
World Laboratory for Cosmology and Particle Physics (WLCAPP), Cairo, Egypt\\
E-mail: \email{a.tawfik@eng.mti.edu.eg}
}

\abstract{The successive stages of a high-energy collision are conjectured to end up with chemical and thermal freezeout of the produced particles. We utilize generic (non)extensive statistics which is believed to determine the degree of (non)extensivity through two critical exponents due to possible phase-space modifications. This statistical approach likely manifests various types of correlations and fluctuations and also possible interactions among the final-state produced particles. We study the baryon-to-pion ratios at top RHIC and LHC energies including the so-called proton anomaly.}

\FullConference{38th International Conference on High Energy Physics\\
		 3-10 August 2016\\
		Chicago, USA}

\begin{document}

\section{Introduction}
\label{Intro}

In statistical-thermal models, in which an {\it ad hoc} extensivity is assumed, i.e. Boltzman-Gibbs (BG), statistical fits have been successfully utilized in order to describe the final-state produced particles such as multiplicity, rapidity and transverse momentum distributions and yields and ratios of specific particle species observed at a wide range of collision energies \cite{Tawfik:2014eba}. A clear picture about the nuclear phase-diagram can be drawn from the variation of the chemical freezeout temperature with the baryon chemical potential or the collision energies. It was surprisingly observed that at top RHIC and LHC energies, the measured proton-to-pion ratios \cite{Abelev:2012} seem not agreeing with the {\it extensive} statistical-thermal models \cite{Becattini:2013}. A possible explanation suggests that the freezeouts occur in chemical and thermal out-of-equilibrium \cite{Petran:2013}, which could be taken into consideration through ingredients such as excluded volume corrections and nonequilibrium occupation factors, etc. added to the extensive statistics. 

Here, we assume that both equilibrium and nonequilibrium in the particle production are best simulated by generic (non)extensive statistical approach. Tsallis statistics \cite{Tsallis1988} is a well-known example on nonextinsivity. It enters an additional parameter, $q$, and leads to very low freezeout temperatures relative to the ones deduced from the BG fits of the particle ratios and yields  and the transverse momentum distributions \cite{Tawfik:2016pwz}. 
The generic (non)extensivite statistical approach determines the degree of (non)extensivity through an equivalence class $(c,d)$, to which a scaling function is assigned. This is characterized by the exponent $c$ or $d$ for first or second property, respectively. The well-know BG extensivity and Tsallis nonextensivity  are retrieved at $(1,1)$ and $(q,0)$, respectively. Avoiding the {\it ad hoc} implementation of either BG or Tsallis to the particle production, as they are just very special cases in the $(c,d)$-space, is a great advantage of the proposed genetic (non)extensivity, especially that no theory is available so far describing the process of the particle production. Secondly, the proposed approach assumes that the possible modifications in the phase space likely autonomously determine whether the system of interest is to be characterized by extensive or nonextensive statistics.

\section{Generic (non)extensive statistics}
\label{sec:model}

Generalized entropy is well classified according to their asymptotic properties, which are categorized into an equivalent class $(c,d)$ \cite{Thurner1,Thurner2,Tawfik:2016pwz}
\begin{eqnarray}
S_{c,d}[p] &=& \sum_{i=1}^{N} {\cal A}\, \Gamma(d+1, 1 - c \log p_i) - {\cal B}\, p_i, \label{eq:NewExtns1}
\end{eqnarray}
where $N$ is the number of micro-states or processes taking place in a complex system and  $\Gamma (a, b)=\int_{b}^{\infty}\, dt\, t^{a-1} \exp (-t)$ is incomplete gamma-function. ${\cal A}$ and ${\cal B}$ are arbitrary parameters. The scaling exponents $(c, d)$ not only characterize both extensive and nonextensive entropies, but also specifies the correspondent generalized exponential and logarithmic functions, respectively \cite{Thurner1},
\begin{eqnarray}
\varepsilon_{c,d,r}(x) &=& \exp\left\{\frac{-d}{1-c}\left[\mathtt{W}_k\left(B\left(1-x/r\right)^{1/d}\right) - \mathtt{W}_k(B)\right]\right\}, \hspace*{3mm} \label{eq:ps1} \\
\Lambda_{c,d,r}(x) &=& r \left\{1-x^{c-1}\left[1- \frac{1-r(1-c)}{d r } \log(x)\right]^d\right\}, \label{eq:genLog}
\end{eqnarray}
where $\mathtt{W}_k$ is the $k$-th Lambert-{$\mathtt{W}$} function and $B \equiv  (1-c)r[1-(1-c)r]\, \exp \left\{(1-c)r/[1-(1-c)r]\right\}$, with $r=(1-c+c d)^{-1}$. 
The scaling exponents can be parametrized as \cite{Thurner1}
\begin{eqnarray}
\left(1-c\right)^{-1} = \lim_{N \rightarrow \infty} N\, \frac{d}{d N}\, \log W(N), & \; &
d = \lim_{N \rightarrow \infty} \log W(N)\, \left[c-1 + \left(N\, \frac{d}{d N}\, \log W(N)\right)^{-1}\right], \label{6}
\end{eqnarray}
where $W(N)$ gives the number of states in a system composed of $N$ micro-states \cite{Thurner1,Thurner2}
\begin{eqnarray}
W(N) &=& \frac{1}{\varepsilon_{c,d}(-\varphi\, c\, N)} 
  \exp \left\{\frac{d}{1-c} \mathtt{W}_k \left[\frac{(1-c)\, e^{\frac{1-c}{c\, d}}}{c\, d} \left(\frac{\varphi\, c\, N}{r} \right)^{1/d}\right] \right\}, \label{eq:states1}
\end{eqnarray}
and $\varphi$ is determined from the generalized entropy $\varphi = d S_g/d N$.

In classical hadron gas composed of $N$ resonances and from Eq. (\ref{eq:ps1}), the partition function at finite temperature ($T$) and chemical potential ($\mu$) is given as
\begin{eqnarray}
\ln\, Z_{cl}(T, \mu) &=& V\, \sum_i^{N}\, g_i \int_0^{\infty} \frac{d^3 p}{(2 \pi)^3}\; \varepsilon_{c,d,r}(x_i), \label{eq:PFcdr1}
\end{eqnarray}
where $x_i=[\mu_i-(p^2+m_i^2)^{1/2}]/T$, $g_i$ being $i$-the resonance's dispersion relation and degeneracy factor and $V$ is the fireball volume. For Fermi-Dirac and Bose-Einstein quantum statistics, 
\begin{eqnarray}
\ln\, Z_{M|B}(T, \mu) &=& \pm V\, \sum_i^{N}\, g_i \int_0^{\infty} \frac{d^3 p}{(2 \pi)^3}\; \Lambda_{c,d,r}\left(1\pm\varepsilon_{c,d,r}(x_i)\right),  \hspace*{4mm} \label{eq:PFcdr2}
\end{eqnarray}
where $\pm$ stands fermions and bosons, respectively. All thermodynamic quantities can be derived, 
\begin{eqnarray}
p(T, \mu) = \frac{\partial}{\partial V}\, \ln\, Z(T,\mu), \qquad & & \qquad 
n(T, \mu) = \frac{\partial}{\partial \mu}\,  p(T,\mu). \label{eq:n}
\end{eqnarray}

At the stage of chemical freezeout, the number of produced particles is conjectured to be fixed and the produced resonances are assumed to complete their decays either to stable particles or to other resonances. Accordingly, the contributions of $i$-th stable particle or resonance to the number density, Eq. (\ref{eq:n}), for instance, have to take into account both possibilities  \cite{Tawfik:2006dk}
\begin{eqnarray}
\langle n_{i}^{\mathtt{final}}(T, \mu)\rangle &=& \langle n_{i}^{\mathtt{direct}}(T, \mu)\rangle  +\sum_{j\neq i} b_{j\rightarrow i} n_{j}(T, \mu), \label{eq:n2}
\end{eqnarray}
where $b_{j\rightarrow i}$ is the branching ratio of $j$-th resonance decaying into $i$-th stable particle or resonance. Great details about the hadron resonance gas (HRG) model can be taken from the recent review article \cite{Tawfik:2014eba}, in which extensive statistics was assumed, exclusively. Also, more details about the proposed generic (non)extensive approach can be found in Ref. \cite{Tawfik:2016pwz}.

\section{Results and conclusions}

\begin{figure}[htb]
\centering{
\includegraphics[width=5.5cm,angle=-0]{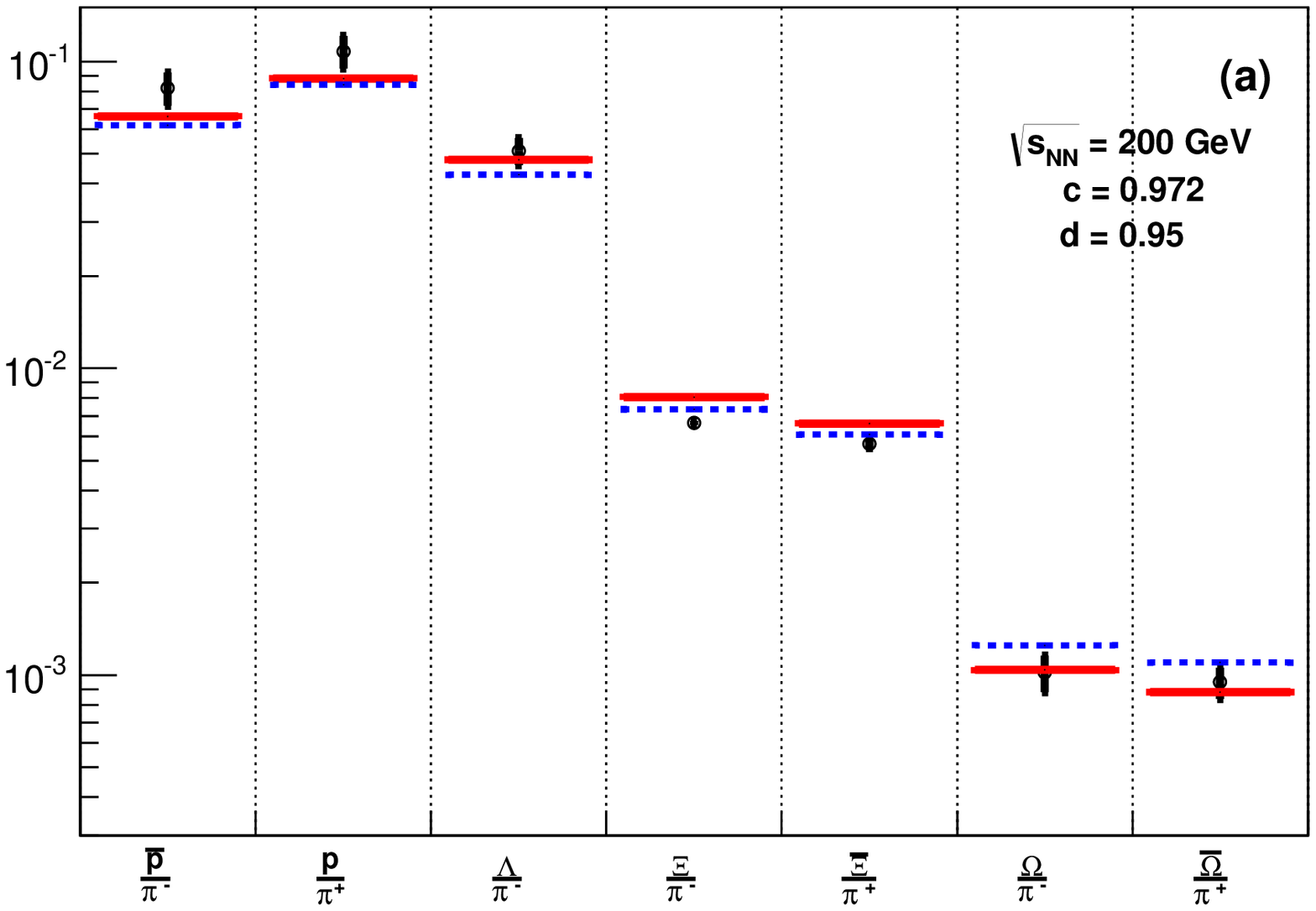}
\includegraphics[width=5.5cm,angle=-0]{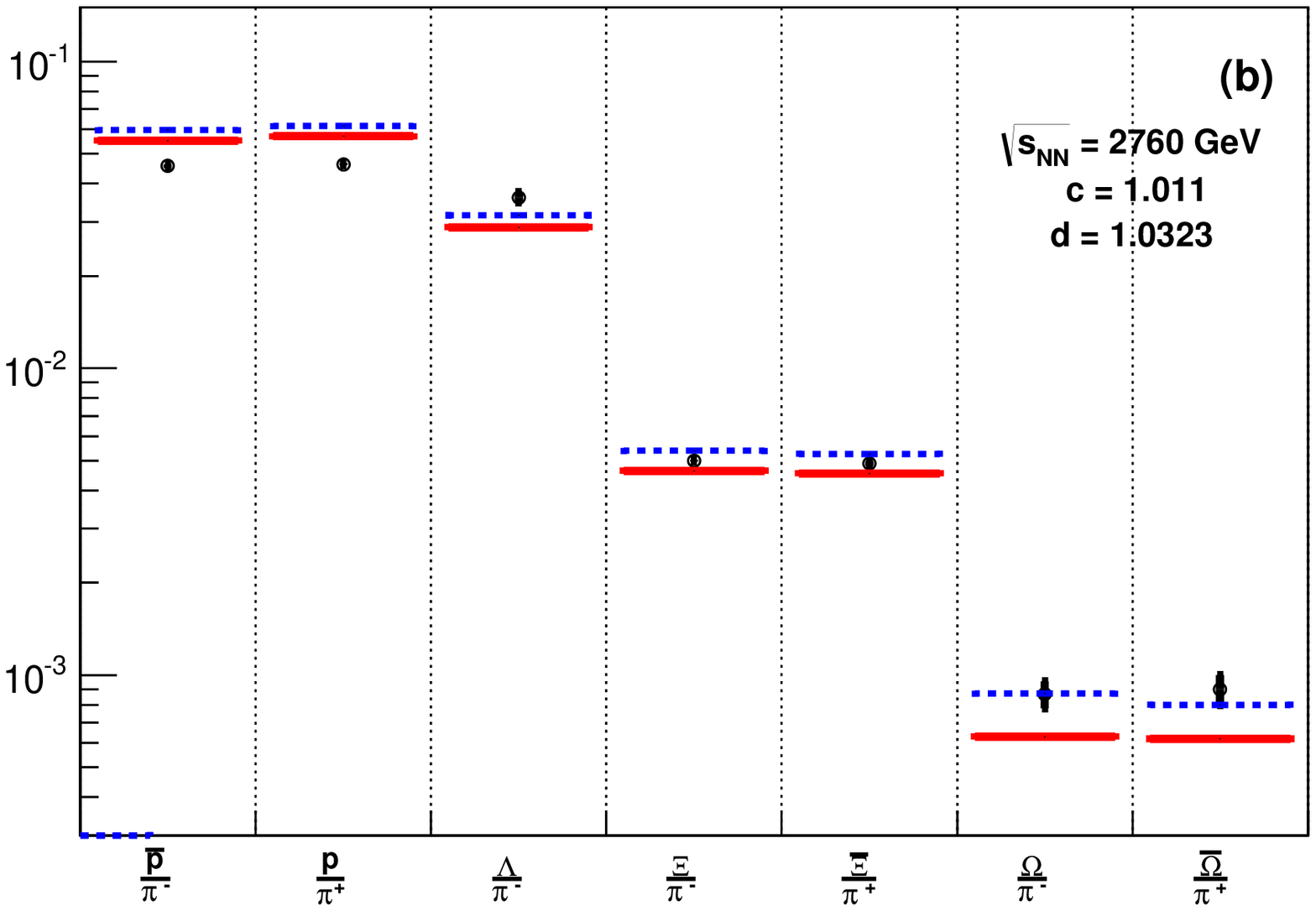}
\caption{\footnotesize Comparison of the results from our nonextensive HRG model for various baryon-to-pion ratios (lines) with experimental measurements (open symbols with errors) at RHIC (left-hand panel) and LHC (right-hand panel). \label{fig:1}
}
}
\end{figure}

\begin{table}[htb]
\begin{center}
{\small 
\begin{tabular}{||c||c|c|c|c|c|c||}
\hline \hline  
 & $\sqrt{s_{NN}}$ GeV & $T$ MeV  & $\mu$ MeV & $c$ & $d$ & $\chi^2$/dof \\ 
\hline \hline 
Extensive & $200$ & $160.65$ & $24.95$& - & - & $1.51$ \\ 
              & $2760$ & $153.09$ & $2.52$& - & - & $3.16$ \\ 
\hline 
Nonextensive & $200$ & $156.24$ & $22.30$& $0.972$ &$0.95$ &  $1.35$ \\ 
              & $2760$ & $161.71$ & $2.52$& $1.011$ &$1.0323$ & $3.18$ \\ 
\hline \hline               
\end{tabular} 
\caption{Nonextensive and extensive fit parameters.}
\label{tab:1}
}
\end{center}
\end{table}

The baryon-to-pion ratios, $\bar{\mathrm{p}}/\mathrm{\pi}^-$, $\mathrm{p}/\pi^+$, $\mathrm{\Lambda}/\pi^-$, $\mathrm{\Xi}/\pi^-$, $\bar{\mathrm{\Xi}}/\pi^+$, $\mathrm{\Omega}/\pi^-$, and $\bar{\mathrm{\Omega}}/\pi^+$  measured in most central Au$+$Au collisions at $\sqrt{s_{NN}}=200~$GeV (open symbols with errors)  are fitted by HRG with generic nonextensive statistics (solid lines). These are compared with extensive HRG (dashed lines) \cite{Tawfik:2014eba}. Both results are depicted in the left-hand panel of Fig. \ref{fig:1}. The thermal fits of the same set of particle ratios measured in most central Pb$+$Pb collisions at $\sqrt{s_{NN}}=2760~$GeV to extensive and generic-nonextensive HRG calculations are presented in the right-hand panel. The resulting nonextensive and extensive fit parameters at both collision energies are listed in Tab. \ref{tab:1}.

We observe that the various baryon-to-pion ratios are well reproduced by the HRG model with generic nonextensive statistics. Furthermore, we notice that the resulting freezeout parameters are very compatible with each others. This means that the (non)extensivity is not necessarily related to radical changes in thermodynamic quantities, such as temperature, but rather to the equivalent class $(c,d)$. Last but not least, the resulting $c$ and $d$ mean that the nonextensivity in the particle production is neither BG nor Tsallis. If $c>1$ and $d>1$, the generalized entropy $S_g[p]=\sum_{i=1}^N g(p_i)$ is no longer maximal
for an equi-distribution $p_i=1/N$. If $c<1$ and $d<1$, $S_g[p]$ is given by Lambert-$W_g$ exponentials.




\begin{thebibliography}{99}

\bibitem{Tawfik:2014eba} Abdel Nasser Tawfik, 
Int. J. Mod. Phys. A {\bf 29}, 1430021  (2014). 

\bibitem{Abelev:2012}  B. Abelev {\it et al.} [ALICE Collaboration], Phys. Rev. Lett. {\bf 109}, 252301 (2012).
  
\bibitem{Becattini:2013} F. Becattini {\it et al.}, Phys. Rev. Lett. {\bf 111}, 082302 (2013).

\bibitem{Petran:2013} M. Petran, J. Letessier, V. Petracek and J. Rafelski, Phys. Rev. C {\bf 88}, 034907 (2013).  

\bibitem{Tsallis1988} C. Tsallis, J. Stat. Phys. {\bf 52}, 479 (1988).

\bibitem{Tawfik:2016pwz} Abdel Nasser Tawfik, 
Eur. Phys. J. A {\bf 52}, 253 (2016). 

\bibitem{Thurner1} R. Hanel and S. Thurner, 
Europhys. Lett. {\bf 93}, 20006 (2011).

\bibitem{Thurner2} R. Hanel and S. Thurner, 
Europhys. Lett. {\bf 96}, 50003 (2011).


\bibitem{Tawfik:2006dk} A. Tawfik, 
  Indian J. Phys. {\bf 86}, 642 (2012). 



\end{thebibliography}
\end{document}